\documentstyle[prb,aps,psfig,twocolumn,
]{revtex}
\begin{document}

\onecolumn

\title{Analytical Hartree-Fock gradients with respect to  the cell parameter:
systems periodic in one and two dimensions}
\author{K. Doll} 
\address{Institut f\"ur Mathematische Physik, TU Braunschweig,
Mendelssohnstra{\ss}e 3, D-38106 Braunschweig, Germany}
\author{R. Dovesi, R. Orlando} 
\address{Dipartimento di Chimica IFM, Universit\`a
 di Torino, Via Giuria 5, I-10125 Torino, Italy}

\maketitle

\begin{abstract}
Analytical Hartree-Fock gradients with respect to the cell parameter have
been implemented in the electronic structure code CRYSTAL, for the
case of one and two-dimensional periodicity. As in most molecular codes,
Gaussian type orbitals are used to express the wavefunction.
Examples demonstrate that
the gradients have a good accuracy.
\end{abstract}

\vspace{1cm}

Keywords: Hartree-Fock, cell gradient, periodic systems, CRYSTAL

\pacs{ }

\narrowtext
\section{Introduction}

Analytical gradients 
\cite{PulayAdv,PulayChapter,Helgaker,HelgakerJorgensen1992,SchlegelYarkony,PulayYarkony,Schlegel2000}
have become a standard tool in molecular quantum 
chemistry. They are indispensable for the optimization of structures,
and many properties can be efficiently computed with the help
of analytical derivatives.
The field was pioneered by Pulay\cite{Pulay};
the theory had already been derived earlier independently\cite{Bratoz}.

The traditional quantum chemical methods are difficult to apply to solids
because of the large increase of the computational effort with the
system size. After several decades of development, Hartree-Fock
calculations for solids can nowadays be routinely performed with the
CRYSTAL code\cite{CRYSTALbuch,Manual03}. Although Hartree-Fock calculations
often have large errors due to the neglect of electronic correlation,
a large interest has grown in the past few years due to the success
of hybrid functionals which include an admixture of exact (Fock) exchange.

Analytical gradients in the CRYSTAL code were first implemented with
respect to nuclear positions\cite{IJQC,CPCarticle},
and after the implementation of a scheme for geometry optimization, 
an efficient structural optimization could be performed \cite{MimmoArcoetal}.
In periodic systems, the cell parameter is another variable to be optimized.
The first gradients with respect to the cell parameter, 
at the Hartree-Fock level,
were for systems periodic in one dimension\cite{Teramae198384}. 
Various groups have implemented these 
gradients in one dimension\cite{Jacquemin,Hirata1997} (see also
the recent review article\cite{Champagne2005}) or
in two dimensions\cite{Tobita2003}. For the general case, a strategy
to compute cell parameter derivatives (and thus the stress tensor)
was suggested with point charges \cite{Kudin2000}, and
an algorithm for structural optimization, based on 
redundant internal coordinates was proposed\cite{Kudin2001}.
Second analytical derivatives with respect to the cell parameter
have also been implemented recently\cite{Jacquemin2003}.

A first big step of the corresponding 
implementation in the CRYSTAL code were analytical 
Hartree-Fock gradients with respect to the cell parameter
in three dimensions \cite{TCA2004}. It is important to note
that the CRYSTAL code is based on the Ewald\cite{Ewald,VicCoulomb} 
method in three dimensions,
so that computing analytical gradients with respect to the cell
parameter requires various additional derivatives: for example the
reciprocal lattice vectors depend on the cell parameter, and various
others. This requires additional derivatives which were not yet
available with the implementation of nuclear gradients, and this
has been documented in great detail \cite{TCA2004}. The
one and two-dimensional case are again different because different potentials
are used: Parry's potential in two dimensions\cite{Parry,Heyes}, and
Saunders' potential in one dimension\cite{Vic1994}. Parry's potential
is similar to Ewald's potential, but modified for the case of two dimensions.
Saunders' potential relies on a real space approach.

This article is intended to complement the first article on 
cell gradients\cite{TCA2004}. Many parts have already been described
in the first article, and therefore the main emphasis is to 
delineate the differences due to the dimensionality.
The article consists thus of one section about the general differences
to the three-dimensional case,
one section about the two-dimensional case,
one section about the one-dimensional case, and one section with
examples.

\section{General differences with respect to the three-dimensional case}

The main difference to the three-dimensional case is the way how
the Coulomb energy is computed. The expression to be evaluated is
the Coulomb energy per cell:

\begin{eqnarray} & 
E^{\rm coul}= &
\frac{1}{2}
\int d^3r \int d^3r\: ' \rho(\vec r) \Phi(\vec r - \vec r \: ') 
\rho(\vec r\: ')\nonumber \\
\end{eqnarray}

with $\Phi$ being the potential function corresponding
to three dimensions (Ewald's potential function)\cite{Ewald,VicCoulomb}, 
two dimensions (Parry's potential function) \cite{Parry} or one dimension
(Saunders' potential function) \cite{Vic1994}.

$\rho(\vec r)$ is a cellular charge distribution,
composed of the nuclear charges $Z_a$ at
the positions of the nuclei $\vec A_a$,

\begin{eqnarray}
\rho^{\rm nuc}(\vec r)=\sum_a Z_a \delta(\vec r-\vec A_a)
\end{eqnarray}

and the electronic charge distribution

\begin{eqnarray} & &
\rho^{\rm el}(\vec r)=-\sum_{\vec g,\mu ,\nu}P_{\nu\vec g\mu\vec 0}
\phi_{\mu}(\vec r-\vec A_{\mu})
\phi_{\nu}(\vec r - \vec A_{\nu}-\vec g)
\end{eqnarray}

The basis functions $\phi_{\mu}(\vec r-\vec A_{\mu}-\vec g)$
are real spherical Gaussian type functions, $P_{\nu\vec g\mu\vec 0}$
is the density matrix in real space. 
$\vec A_{\mu}$ denotes the nucleus where the basis function $\mu$
is centered. The implementation is done for the case of closed shell
Hartree-Fock and unrestricted Hartree-Fock methods. For the sake of simplicity,
the spin is ignored in the equations in this article. 
The extension is straightforward,
as was shown for the three-dimensional case\cite{TCA2004}. Examples for
spin-polarized calculations are given in section \ref{Beispielsection}.

The potential function enters via the nuclear-nuclear repulsion
(equation 10 in reference \onlinecite{TCA2004}), the nuclear
attraction integrals (equation 34 in reference \onlinecite{TCA2004}),
and the field integrals (equation 43 in reference \onlinecite{TCA2004}).
Essentially, the derivatives are computed as described in the
previous article \cite{TCA2004}, there are only minor differences
as described in section \ref{2Dsection} and \ref{1Dsection}.

The derivatives of the other integrals (overlap, kinetic energy, 
multipoles, bielectronics) and the calculation of the
energy-weighted density matrix is practically identical to the
three-dimensional case \cite{TCA2004}. 

Finally, the correction due to the
spheropole (equation 47 in reference \onlinecite{TCA2004})
is zero in one and two dimensions and thus does not have
to be discussed. The spheropole is a correction which arises due to
the Ewald method, when applied to the electronic charge distribution:
the charge distribution is approximated by multipoles in the long range, and
not approximated in the short range. The electrostatic potential
is then computed as the sum of the Ewald potential of the multipoles
and of the Coulomb potential of the charge distribution in the 
short range. Replacing the Ewald potential
with the Coulomb potential is correct,
if the difference of multipolar charge distribution 
and the exact charge distribution in the short range,
has zero charge, dipole, quadrupole, and second spherical moment 
\cite{VicCoulomb}.  The second spherical moment can also
be seen as the average electrostatic potential of a charge distribution
(see the discussion in section 3.2 of reference \onlinecite{VicCoulomb}).
Here, it corresponds to the average electrostatic potential of the difference
of the exact and the approximated charge distribution. This term
is finite and in general
non-zero, in the case of periodicity in  three dimensions. 
However, when the system has periodicity in less than  three dimensions, 
the average electrostatic potential of a charge distribution with
zero total charge, dipole and quadrupol, is zero. Therefore, there
is no spheropole in less than three dimensions.

This can also seen from equation 31 in reference \onlinecite{VicCoulomb}.
The average Coulomb potential is obtained as follows:

\begin{eqnarray}
\Phi=-\frac{2\pi}{3V_{cell}}\int \rho^{diff} (\vec r) \ \vec r^2 \ d^3r
\end{eqnarray}

$\rho^{diff}(\vec r)$ 
corresponds here to the difference between the exact charge
distribution and the multipolar charge distribution.
The integral is over the whole space and 
finite. The prefactor involves a division by the cell volume  $V_{cell}$ of the
three-dimensional cell.
We might now approximate a system with periodicity in two dimensions
by a system of slabs with three-dimensional periodicity, where the slabs
are separated by a vacuum region. When we increase the vacuum region
and thus the cell volume $V_{cell}$, then
the integral remains essentially constant, but the prefactor becomes
smaller and smaller and therefore the average Coulomb potential becomes zero,
and the spheropole correction becomes zero. 

It should be mentioned, that two-dimensional periodicity is implemented
in the CRYSTAL code
in such a way that there is only one slab which is not repeated
in the third dimension. Still, the argument presented above holds
in a similar way, and there is thus no spheropole correction in systems
with less than three-dimensional periodicity.

The total energy is thus
similar to the three-dimensional case\cite{TCA2004}, 
apart from the spheropole term which is zero:

\begin{eqnarray} & & 
E^{\rm total}=E^{\rm kinetic}+E^{\rm NN}+E^{\rm coul-nuc}+E^{\rm coul-el}
+E^{\rm exch-el}=\nonumber \\ & & 
=\sum_{\vec g,\mu,\nu}P_{\nu\vec g\mu\vec 0}T_{\mu\vec 0\nu\vec g}+
\frac{1}{2}\sum_{a,b} Z_a Z_b \Phi(\vec A_{b}-\vec A_{a})
\nonumber \\ & & 
-\sum_{\vec g,\mu,\nu} 
P_{\nu\vec g\mu\vec 0}\sum_{a}Z_a\int
\phi_{\mu}(\vec r-\vec A_{\mu})
\phi_{\nu}(\vec r-\vec A_{\nu}-\vec g)
\Phi(\vec r-\vec A_{a}){\rm d^3r}
\nonumber \\ & & 
+\frac{1}{2}\sum_{\vec g,\mu,\nu} P_{\nu\vec g\mu\vec 0}
\bigg(\sum_{\vec h,\tau,\sigma}
P_{\sigma\vec h\tau\vec 0}
C_{\mu\vec 0\nu\vec g \tau\vec 0\sigma\vec h}
-\sum_c \sum_{l=0}^{L}\sum_{m=-l}^{l}\eta_l^m(\rho_c;\vec A_c)
M_{l\mu\vec 0\nu\vec gc}^m\bigg)\nonumber \\ & & 
-\frac{1}{2}\sum_{\vec g,\mu,\nu}P_{\nu\vec g\mu\vec 0}
\sum_{\vec h,\tau,\sigma}P_{\sigma\vec h\tau\vec 0}
X_{\mu\vec 0\nu\vec g\tau\vec 0\sigma\vec h}
\end{eqnarray}

The individual terms contributing to the total energy are the kinetic energy 
$E^{\rm kinetic}$,
the nuclear-nuclear repulsion energy $E^{\rm NN}$, the nuclear-electron
attraction $E^{\rm coul-nuc}$, the electron-electron repulsion
$E^{\rm coul-el}$ and the Fock exchange $E^{\rm exch-el}$. The variables
will not all be explained in order to reduce the number of formulas in this
article. The reader is referred to the article on the three-dimensional
case for the details where all these terms are explained\cite{TCA2004}.
The gradient with respect to the cell parameters $a_{ij}$
is given in the following equation. As the total energy, the gradient
is similar to the
three-dimensional case apart from the spheropole term which is zero.

\begin{eqnarray}
\label{Forcecellparametergleichung} & &
F_{a_{ij}}=-\frac{\partial E^{\rm total}}{\partial a_{ij}}=\nonumber \\ & &
-\sum_{\vec g,\mu,\nu}P_{\nu\vec g\mu\vec 0}\frac{\partial 
T_{\mu\vec 0\nu\vec g}}
{\partial a_{ij}}
-\frac{\partial E^{\rm NN}}{\partial a_{ij}}
-\sum_{\vec g,\mu,\nu} P_{\nu\vec g\mu\vec 0}\frac{\partial 
N_{\mu\vec0\nu\vec g}}{\partial a_{ij}}
\nonumber \\ & &
-\frac{1}{2}\sum_{\vec g,\mu,\nu} P_{\nu\vec g\mu\vec 0}
\bigg\{
\sum_{\tau,\sigma}P_{\sigma\vec h\tau\vec 0}
\frac{\partial C_{\mu\vec 0\nu\vec g \tau\vec 0\sigma\vec h}}
{\partial a_{ij}}
-\sum_c \sum_{l=0}^{L}\sum_{m=-l}^{l}\sum_{\vec h,\tau \in  c, \sigma}
P_{\sigma\vec h \tau \vec 0} \nonumber \\ & & 
\frac{\partial}{\partial a_{ij}}\bigg[\int
\phi_{\tau}(\vec r-\vec A_{\tau})
\phi_{\sigma}(\vec r - \vec A_{\sigma}-\vec h)
X_l^m(\vec r-\vec {A_c}){\rm d^3r} \
 M_{l\mu\vec 0\nu\vec g c}^m\bigg]\bigg\}
\nonumber \\ & &
+\frac{1}{2}\sum_{\vec g,\mu,\nu} P_{\nu\vec g\mu\vec 0}
\sum_{\vec h,\tau,\sigma}P_{\sigma\vec h\tau\vec 0}
\frac{\partial X_{\mu\vec 0\nu\vec g\tau\vec 0\sigma\vec h}}
{\partial a_{ij}} 
\nonumber \\ & &
-\sum_{\vec g,\mu,\nu}
\frac{\partial S_{\mu\vec 0\nu\vec g}}{\partial a_{ij}}  \int_{BZ} 
 \exp({\rm i}\vec K\vec g)\sum_n
a_{\nu n}(\vec K)a_{\mu n}(\vec K)
\epsilon_n(\vec K)
\Theta(\epsilon_F-\epsilon_n(\vec K)) {\rm d^3k}
\end{eqnarray}

\section{The two-dimensional case}
\label{2Dsection}

In the two-dimensional case,
the primitive cell is given by two vectors, with two components: 
$\vec a_1$, $\vec a_2$.
$a_{ij}$ are defined in such a way that $a_{11}=a_{1x}$ 
is the
$x$-component of $\vec a_{1}$, $a_{12}=a_{1y}$ the $y$-component of $\vec a_1$,
$a_{21}$ is the $x$-component of $\vec a_{2}$, and $a_{22}$ is the
$y$-component of $\vec a_{2}$.

\begin{eqnarray}
\left(
\begin{array}{c} 
\vec a_1\\
\vec a_2\\
\end{array}
\right)=
\left(
\begin{array}{c} 
a_{1x} \ a_{1y} \\
a_{2x} \ a_{2y} \\
\end{array}
\right)=
\left(
\begin{array}{c} 
a_{11} \ a_{12} \\
a_{21} \ a_{22} \\
\end{array}
\right)
\end{eqnarray}

A point $\vec g$ of the direct lattice is defined 
as $\vec g=n_1 \vec a_1+n_2 \vec a_2$, with $n_1, n_2$
being integer numbers. 
The position 
of an atom $c$ in a cell at the origin 
(i.e. $\vec g=\vec 0$) is given as
$\vec A_{c}=f_{c,1}\vec a_1+f_{c,2}\vec a_2$, and then
in cell $\vec g$ the position will be:

$\vec A_{c}+\vec g=(f_{c,1}+n_{\vec g,1})\vec a_1+(f_{c,2}+
n_{\vec g,2})\vec a_2$

We have used an additional index, i.e. 
$n_{\vec g,1}$ means factor $n_1$ of the lattice vector $\vec g$.
The cartesian $t$ component (with $t$ being $x$ or $y$)
of the vector $\vec A_{c}+\vec g$, indicated
as $A_{c,t}+g_{t}$, is thus

$A_{c,t}+g_{t}=\sum_{m=1}^2 (f_{c,m}+n_{\vec g,m})a_{mt}$

As all the integrals depend on the position of the nuclei, the derivatives
of the nuclear coordinates with respect to the cell parameters are
required:

\begin{eqnarray}
\frac{\partial A_{c,t}+g_{t}}{\partial a_{ij}}=
\sum_{m=1}^2 (f_{c,m}+n_{\vec g,m})\delta_{im}\delta_{jt}=
(f_{c,i}+n_{\vec g,i})\delta_{jt}
\end{eqnarray}

with the Kronecker symbol $\delta_{jt}$.

The main difference, compared to the three-dimensional case,
is Parry's potential function $\Phi(\vec r-\vec A_{a})$ that is used:

\begin{eqnarray} &  
\Phi(\vec r-\vec A_{a}) & = \sum_{\vec h}^{'} 
\frac{{1-\rm erf}
(\sqrt\gamma|\vec r-\vec A_{a}-\vec h|)}{|\vec r-\vec A_{a}- \vec h|}
\nonumber \\ & &
+\sum_{\vec K}^{'}\frac{\exp({2\pi{\rm i}
(K_x (x-A_{a,x})+K_y (y-A_{a,y}))})}{2V|\vec K|}
\left(\exp(2\pi|\vec K|(z-A_{a,z})) \left(1-{\rm erf}(\sqrt\gamma(z-A_{a,z})+
\frac{\pi|\vec K|}{\sqrt\gamma})\right) \right. \nonumber \\ & &
\left. +\exp(-2\pi|\vec K|(z-A_{a,z}))\left(1+{\rm erf}
\left(\sqrt\gamma(z-A_{a,z})-
\frac{\pi|\vec K|}{\sqrt\gamma}\right)\right)\right)
\nonumber \\ & &
-\frac{2\pi}{V}(z-A_{a,z}){\rm erf}(\sqrt\gamma(z-A_{a,z}))
-\frac{2\sqrt\pi}{V\sqrt{\gamma}}{\rm exp}{\left(-\gamma(z-A_{a,z})^2\right)}
\end{eqnarray}

$\vec h$ are the direct lattice vectors, $\vec K$ the reciprocal
lattice vectors. $V$ is the area of the two dimensional unit cell, 
$\gamma$ is a screening parameter 
which was optimized to be $\gamma=(2.4/V^{1/2})^2$, in
the two dimensional case. Note that this is different
to the three-dimensional case\cite{VicCoulomb} where $\gamma$ was chosen as 
$\gamma=(2.8/V^{1/3})^2$.
The prime in the direct lattice summation 
indicates that the summation includes all values of the
direct lattice vector $\vec h$, with the exception
of the case when $|\vec r-\vec A_{a}-\vec h|$ vanishes. 
In this case, the term $\frac{1}{|\vec r-\vec A_{a}-\vec h|}$ is omitted
from the sum. In the 
reciprocal lattice series, the prime indicates that
all terms with $\vec K \neq \vec 0$
are included.

The error function $erf$ is defined as in reference \onlinecite{TCA2004},
equation 12.

Like the Ewald potential, Parry's potential depends on the variables
$\vec A_{c}$ , $\vec h$, $V$, $\gamma$ and $\vec K$.
The derivative with respect to the cell parameters thus requires
derivatives with respect to these variables. For the derivatives with
respect to $\vec A_{c}$ and $\vec h$ this is like in the 
three-dimensional case. There are minor changes due to the 
two-dimensionality for the derivatives of the area $V$, of the $\vec K$-vectors with respect to $a_{ij}$ and of the screening parameter $\gamma$.

\subsubsection{Derivative of the area}

The area is obtained as the magnitude of the
cross product of the cell parameters:

\begin{eqnarray}
V=|\vec a_1\times \vec a_2|=|a_{1x}a_{2y}-a_{1y}a_{2x}|
\end{eqnarray}

If we assume that $a_{1x}a_{2y}-a_{1y}a_{2x}$ is positive, then
the derivatives $\frac{\partial V}{\partial a_{ij}}$
are obtained as:

\begin{eqnarray}
\frac{\partial V}{\partial a_{1x}}=a_{2y}\\
\frac{\partial V}{\partial a_{1y}}=-a_{2x}\\
\frac{\partial V}{\partial a_{2x}}=-a_{1y}\\
\frac{\partial V}{\partial a_{2y}}=a_{1x}
\end{eqnarray}

Essentially, the formulas for the three-dimensional case can be used,
when setting $a_{1z}=0, \ a_{2z}=0$ and $\vec a_3=(0,0,1)$. This holds
also for the derivatives of the reciprocal lattice vectors, as described
in the following paragraph.

\subsubsection{Derivative of the reciprocal lattice vectors}

The reciprocal lattice vectors $\vec K$ can be expressed as

\begin{eqnarray}
\vec K=n_1 \vec b_1 + n_2 \vec b_2 
\end{eqnarray}

with the primitive vectors $\vec b_i$ of the reciprocal lattice
defined as:

\begin{eqnarray}
\vec b_1=\frac{2\pi}{V}(a_{2y},-a_{2x}) \ ; \ \
\vec b_2=\frac{2\pi}{V}(-a_{1y},a_{1x})
\end{eqnarray}

The derivatives are thus:

\begin{eqnarray}
\frac{\partial \vec b_1}{\partial a_{1x}}=
-\frac{\partial V}{\partial a_{1x}}\frac{\vec b_1}{V}
\end{eqnarray}

\begin{eqnarray}
\frac{\partial \vec b_1}{\partial a_{1y}}=
-\frac{\partial V}{\partial a_{1y}}\frac{\vec b_1}{V}
\end{eqnarray}

\begin{eqnarray}
\frac{\partial \vec b_1}{\partial a_{2x}}=
-\frac{\partial V}{\partial a_{2x}}\frac{\vec b_1}{V}
+\frac{2\pi}{V}
\left( 0, -1 \right)
\end{eqnarray}

\begin{eqnarray}
\frac{\partial \vec b_1}{\partial a_{2y}}=
-\frac{\partial V}{\partial a_{2y}}\frac{\vec b_1}{V}
+\frac{2\pi}{V}
\left( 1, 0 \right)
\end{eqnarray}

and

\begin{eqnarray}
\frac{\partial \vec b_2}{\partial a_{1x}}=
-\frac{\partial V}{\partial a_{1x}}\frac{\vec b_2}{V}
+\frac{2\pi}{V}
\left( 0, 1 \right)
\end{eqnarray}

\begin{eqnarray}
\frac{\partial \vec b_2}{\partial a_{1y}}=
-\frac{\partial V}{\partial a_{1y}}\frac{\vec b_2}{V}
+\frac{2\pi}{V}
\left( -1, 0 \right)
\end{eqnarray}

\begin{eqnarray}
\frac{\partial \vec b_2}{\partial a_{2x}}=
-\frac{\partial V}{\partial a_{2x}}\frac{\vec b_2}{V}
\end{eqnarray}

\begin{eqnarray}
\frac{\partial \vec b_2}{\partial a_{2y}}=
-\frac{\partial V}{\partial a_{2y}}\frac{\vec b_2}{V}
\end{eqnarray}

\subsubsection{Derivative of the screening parameter}

The derivative is straightforward, like in the three-dimensional case:

\begin{eqnarray}
\frac{\partial \gamma}{\partial a_{ij}}=
\frac{\partial \gamma}{\partial V}\frac{\partial V} {\partial a_{ij}}=
-\frac{\gamma}{V}\frac{\partial V}{\partial a_{ij}}
\end{eqnarray}

As a whole, Parry's potential leeds to similar terms appearing in the
derivatives as in the case of the Ewald potential. This is what was 
to be expected, as Parry's potential is essentially obtained
when Ewald's approach to treat the Coulomb interaction is applied 
to a system with two-dimensional periodicity.

\section{The one-dimensional case}
\label{1Dsection}

In the one-dimensional case, there is only one cell parameter: 
$a_{xx}=a_{11}=a=|\vec a|$. This case is somewhat different from the two- 
and three-dimensional case because a pure real space approach is used
in the CRYSTAL code
for the potential to describe the Coulomb interaction\cite{Vic1994}.
The potential consists of a point charge $+1$, neutralized by a uniform
charge distribution of length $a$, with charge density $-\frac{1}{a}$.
The uniform charge distribution is then again compensated. Up to 
a certain range, the summation is performed exactly. For larger
distances, the summation is instead approximated with the help of
the Euler-MacLaurin summation rule.
As a whole, the following expression was obtained\cite{Vic1994}:

\begin{eqnarray}
\Phi(\vec r)=\sum_{n=-M}^{M} \prime \frac{1}{|\vec r-n\vec a|}
-\frac{H(U-z,\alpha)+H(U+z,\alpha)}{a}
+\xi(M,\vec r)+\xi(M,-\vec r)
\end{eqnarray}

The first term comprises the exact part, the next two (with the $H$
function) the region due to the uniform charge density in the
range of the exact sum
(from $-M\vec a$ to $M\vec a$), the remaining two
terms (the $\xi$-function) are the approximated part.  
The prime indicates that terms
with $|\vec r-n\vec a|=0$ are omitted. 
$M$ is thus the number of cells, where the sum is performed exactly,
and $U=a(M+\frac{1}{2})$.
$\alpha$ is defined as $\alpha=x^2+y^2$, with $\vec r=(x,y,z)$.
$H$ is the function $H(p,\alpha)=\ln (\sqrt{p^2+\alpha}+p)$. 
$\xi(M,\vec r)$ and $\xi(M,-\vec r)$ are contributions from the long
range part, which is approximated by the Euler-MacLaurin rectange rule
summation formula. For more details,
see reference \onlinecite{Vic1994}. For the present purpose,
it is important to note that the direct lattice vector $a$ appears
in the potential, but no screening parameter $\gamma$ and no
reciprocal lattice vectors $\vec K$ as in the two- and three-dimensional
case. This means that derivatives with respect to the nuclear
coordinates $\vec A_c$ and derivatives with respect to the
direct lattice vectors $n\vec a$ appear,
which are essentially given by the nuclear gradients, multiplied with
the fractional coordinates. The derivatives with respect to $a$ due
to the $H$ and $\xi$ function are very lengthy, but still straightforward.
They are thus not discussed here, but formulas can be derived from
Saunders' article \cite{Vic1994}.

\section{Examples}
\label{Beispielsection}

In this section, we give some numerical examples of the accuracy of
the gradients. The tests considered 
are essentially identical or similar to the test cases
distributed with the CRYSTAL code and with the ones from
reference \onlinecite{MimmoArcoetal}. 
Note that the fractional coordinates of the atoms were not optimized.

First, two systems with one-dimensional periodicity are considered.
In table \ref{SNtable}, SN is
periodically arranged. The analytical and numerical derivative agree
well up to 4 digits, and the minimum of the energy at $a$=4.42 \AA \
agrees with the place where the gradient changes its sign.
In table \ref{Polyglycinetable}, such a comparison is done for
polyglycine. The agreement of numerical and analytical gradients 
is similar to SN, and again the vanishing of the gradient agrees
with the minimum of the energy, to at least 0.01 \AA. In table
\ref{NiOtable}, ferromagnetic NiO is studied at the level of unrestricted
Hartree-Fock. The agreement of numerical and analytical gradient can 
be improved by increasing the "ITOL"-parameters\cite{Manual03}, as
described earlier\cite{IJQC,TCA2004}.
Indeed, when increasing
them from default values to higher ones, symmetric in ITOL4 and ITOL5,
then analytical and numerical gradient match better. Note that, when
running at lower ITOL parameters, an inaccuracy is introduced in the total
energy expression and thus in the numerical gradients as well. The
fact that numerical and analytical gradients match less well at low
ITOL values is thus a combination of an inaccuracy in the energy expression
(which affects the numerical gradient) and an inaccuracy in the analytical
gradient. Still, in all the tests performed so far, no severe error
was found when using default values for the ITOL parameters. Using
higher ITOL parameters is mainly useful for tests of the correctness
of the code.

Then, various systems with two-dimensional periodicity are considered.
In table \ref{MgOtable}, 3 MgO layers are considered. Numerical and
analytical derivative agree to 3 digits, and the minimum of the
energy and the vanishing of the gradient agree also well.
The same accuracy is found 
for Al$_2$O$_3$ in table \ref{Al2O3table}, where a slab 
with 6 atomic layers is considered. 
In table \ref{Cr2O3table}, a Cr$_2$O$_3$ slab was chosen as an example
for unrestricted Hartree-Fock. The accuracy is slightly worse when
comparing the numerical and the analytical gradient. This can
again be improved by increasing the "ITOL"-parameters.
The minimum in the energy
agrees already with default "ITOL" values to at least 0.01 \AA.
Finally, in table \ref{LiFtable}, LiF was arranged with two dimensional
periodicity, without symmetry, in such a way that three components
of the cell gradient ($a_{1x}$, $a_{1y}$, $a_{2y}$) can be computed
independently. This test thus demonstrates that these components 
are correctly computed.

In table \ref{CPUtable}, the CPU times are displayed. The
calculations were performed on a single CPU of a Compaq ES45, with a
clock rate of 1 GHz. As in the three-dimensional case, we 
compare again the CPU time
for the integrals with the time for the gradients.
The CPU time for all the gradients
(nuclear and cell gradients) is roughly five to ten times the CPU time
for the integrals. This may become smaller in the future with
further optimizations in the gradient code. Note that the CPU time
for the self consistent field calculations is relatively high because
a very low convergence threshold was chosen in  order to ensure the
accuracy of the succeeding gradient calculation (the gradient calculation
is the more accurate, the more accurately the self consistent field equations
are solved). 

The CPU times thus indicate that analytical gradients can be
computed at a relatively low expense. Compared with numerical gradients,
it appears that analytical gradients should usually be favorable,
especially because numerical gradients will depend on the step size,
and often it will be necessary to break a symmetry for a finite displacement,
to compute the numerical gradient. Numerical gradients require at least
one additional energy evaluation for each coordinate to be optimized,
which makes analytical gradients clearly favorable, if there is a large number
of geometrical parameters.

\section{Conclusion}

A formalism for the calculation of the analytical gradient 
of the Hartree-Fock energy, with respect
to the cell parameter, has been presented and implemented in the code
CRYSTAL, for the case of systems periodic
in one and two dimensions. The implementation
includes the cases of spin-restricted and unrestricted polarization.
It was
shown that a high accuracy can be achieved. 

\section{Acknowledgment}
The calculations were performed on a Compaq ES45
(computer center of the TU Braunschweig).

\newpage
\onecolumn

\begin{table}
\begin{center}
\caption{SN, with one-dimensional periodicity. A comparison of analytical and
numerical gradient is done for various unit cell lengths. A $[3s2p1d]$
basis set was used for S, and a $[2s1p]$ basis set for N.}
\label{SNtable}
\begin{tabular}{cccc} 
 $a$ & analytical derivative  & numerical derivative & energy \\
$[$\AA] & $[E_h/a_0]$ & $[E_h/a_0]$ & $[E_h]$ \\
4.30  & 0.04144 & 0.0414  &  -893.870081 \\
4.41  & 0.00372 & 0.0037   & -893.874639 \\
4.42  & 0.00064 & 0.0006  &  -893.874680 \\
4.43  & -0.00238 & -0.0024 & -893.874663 \\
4.500 & -0.02208 & -0.0221 & -893.873013\\
\end{tabular}
\end{center}
\end{table}

\begin{table}
\begin{center}
\caption{Polyglycine. A comparison of analytical and
numerical gradient is done for various unit cell lengths. Basis sets
of the size $[2s1p]$ were used for C, O, N and a $[1s]$ basis set for
H.}
\label{Polyglycinetable}
\begin{tabular}{cccc} 
 $a$  & analytical derivative  & numerical derivative & energy \\
$[$\AA] & $[E_h/a_0]$ & $[E_h/a_0]$ & $[E_h]$ \\
7.30 &  0.01956 &  0.0196 & -408.220173 \\
7.42 &  0.00116 &  0.0012 & -408.222495  \\
7.43 & -0.00030 & -0.0003 &  -408.222503 \\
7.44 & -0.00175 & -0.0017 &  -408.222484  \\
7.50 & -0.01018 & -0.0102 & -408.221807\\
\end{tabular}
\end{center}
\end{table}

\begin{table}
\begin{center}
\caption{NiO, ferromagnetic, unrestricted Hartree-Fock. 
The gradient with respect to the cell
parameter is computed for two different values of the ITOL parameters.
A $[5s4p2d]$ basis set for Ni was used, and a $[4s3p]$ basis set for O.}
\label{NiOtable}
\begin{tabular}{cccc} 
 $a$  & analytical derivative  & numerical derivative & energy \\
$[$\AA] & $[E_h/a_0]$ & $[E_h/a_0]$ & $[E_h]$ \\
\multicolumn{4}{c}{ITOL 6 6 6 6 12 (default)} \\
5.00 & -0.10864 & -0.1074 & -1581.454974 \\
\multicolumn{4}{c}{ITOL 6 6 6 12 12} \\
5.00 & -0.10782 & -0.1078 & -1581.456358 \\
\end{tabular}
\end{center}
\end{table}

\begin{table}
\begin{center}
\caption{MgO surface, 3 atomic layers. 
The unit cell consists of 3 Mg and 3
O atoms, with $a_{1x}=a_{2y}=a$. 
Basis sets of the size $[3s2p]$ were used. The derivative 
with respect to $\displaystyle{\frac{\partial}{\partial a}=
\frac{\partial}{\partial a_{1x}}\frac{\partial a_{1x}}{\partial a}+
\frac{\partial}{\partial a_{2y}}\frac{\partial a_{2y}}{\partial a}}$
is displayed, $\displaystyle{\frac{\partial}{\partial a_{1y}}}$ and
$\displaystyle{\frac{\partial}{\partial a_{2x}}}$ do not contribute.}
\label{MgOtable}
\begin{tabular}{cccc} 
 $a$ & analytical derivative  & numerical derivative & energy \\
 $[$\AA] & $[E_h/a_0]$ & $[E_h/a_0]$ & $[E_h]$ \\
2.80 &  0.10544 &  0.1058 & -823.930493 \\
2.88 &  0.01035 &  0.0108 & -823.939034 \\
2.89 &  0.00006 &  0.0006 & -823.939142 \\
2.90 & -0.00991 & -0.0095 & -823.939058  \\
3.00 & -0.09403 & -0.0937 & -823.928906 \\
\end{tabular}
\end{center}
\end{table}

\begin{table}
\begin{center}
\caption{Al$_2$O$_3$, 6 atomic layers. The unit cell consists of 6 Al  and 4 
O atoms, with $a_{1x}=\sqrt{3}/2*a_{2y}=\sqrt{3}/2*a$. 
Basis sets of the size $[3s2p1d]$ for Al and $[2s1p]$ for
O were chosen. The derivative 
with respect to $\displaystyle{\frac{\partial}{\partial a}=
\frac{\partial}{\partial a_{1x}}\frac{\partial a_{1x}}{\partial a}+
\frac{\partial}{\partial a_{2y}}\frac{\partial a_{2y}}{\partial a}}$
is displayed, $\displaystyle{\frac{\partial}{\partial a_{1y}}}$ and
$\displaystyle{\frac{\partial}{\partial a_{2x}}}$ do not contribute.}
\label{Al2O3table}
\begin{tabular}{ccccc} 
 $a$  & analytical derivative  & numerical derivative & energy \\
$[$\AA] & $[E_h/a_0]$ & $[E_h/a_0]$ & $[E_h]$ \\
4.20 &  0.27548 &  0.2757 & -1400.244182 \\
4.40 &  0.00590 &  0.0059 & -1400.295000 \\
4.41 & -0.00570 & -0.0060 & -1400.295003 \\
4.42 & -0.01712 & -0.0171 & -1400.294787 \\
4.70 & -0.27847 & -0.2786 & -1400.211859 \\
\end{tabular}
\end{center}
\end{table}

\begin{table}
\begin{center}
\caption{Cr$_2$O$_3$, 6 atomic layers,  ferromagnetic,
unrestricted Hartree-Fock. The unit cell 
consists of 6 Cr and 4
O atoms, with $a_{1x}=\sqrt{3}/2*a_{2y}=\sqrt{3}/2*a$. 
Basis sets of the size $[5s4p2d]$ for Cr and $[3s2p]$ for O
were chosen. The derivative 
with respect to $\displaystyle{\frac{\partial}{\partial a}=
\frac{\partial}{\partial a_{1x}}\frac{\partial a_{1x}}{\partial a}+
\frac{\partial}{\partial a_{2y}}\frac{\partial a_{2y}}{\partial a}}$
is displayed, $\displaystyle{\frac{\partial}{\partial a_{1y}}}$ and
$\displaystyle{\frac{\partial}{\partial a_{2x}}}$ do not contribute.}
\label{Cr2O3table}
\begin{tabular}{ccccc} 
 $a$  & analytical derivative  & numerical derivative & energy \\
$[$\AA] & $[E_h/a_0]$ & $[E_h/a_0]$ & $[E_h]$ \\
\multicolumn{4}{c}{ITOL 6 6 6 6 12 (default)} \\
4.70 &  0.13465  &  0.1379 & -4622.589785 \\
4.87 &  0.00426  &  0.0069 & -4622.612278 \\
4.88 & -0.00253  &  0.0001 & -4622.612339 \\
4.89 & -0.00921  & -0.0066 & -4622.612277 \\
5.00 & -0.07676  & -0.0745 & -4622.603638 \\
\multicolumn{4}{c}{ITOL 6 6 6 12 12} \\
4.88 & -0.00116  & -0.0011 & -4622.617935 \\
5.00 & -0.07539  & -0.0754 & -4622.609006 \\
\end{tabular}
\end{center}
\end{table}

\begin{table}
\begin{center}
\caption{LiF, with a unit cell of $a_{1x}=5$ \AA, $a_{2y}=4$ \AA, 
and an angle of 60$^\circ$, resulting in $a_{1y}=2.5$ \AA.
The F atoms are at 
(x=0.1, y=0 (x and y in fractional units), z=0.1 \AA),
(x=0.5, y=0.5 (x and y in fractional units), z=0.3 \AA), the Li atoms at
(x=0.5, y=0 (x and y in fractional units), z=0.2 \AA), and 
(x=0, y=0.5 (x and y in fractional units), z=0.4 \AA). 
A $[2s1p]$ basis set was used for Li, a $[4s3p]$ basis set for F.}
\label{LiFtable}
\begin{tabular}{ccccc} 
component & analytical derivative  & numerical derivative \\
 & $[E_h/a_0]$ & $[E_h/a_0]$  \\
$\displaystyle{\frac{\partial E} {\partial a_{1x}}}$ &  0.04045 &  0.0406 \\ \\
$\displaystyle{\frac{\partial E} {\partial a_{1y}}}$ & -0.04415 & -0.0441 \\ \\
$\displaystyle{\frac{\partial E} {\partial a_{2y}}}$ & -0.01838 & -0.0183 \\ \\
\end{tabular}
\end{center}
\end{table}

\begin{table}
\begin{center}
\caption{CPU times for one single point calculation of the various systems. 
The calculations were performed on a Compaq ES45, 
using a single CPU (1 GHz). The CPU times refer to the part for the integrals
(all the integrals were written to disk), the self-consistent field (SCF)
procedure, and to the calculation of all the gradients (i.e. nuclear 
gradients and cell gradients).}
\label{CPUtable}
\begin{tabular}{cccccc} 
system & \multicolumn{3}{c}{CPU time, in seconds} \\
 & integrals & SCF & gradients \\
SN &  1 & 1 & 6\\
Polyglycine & 2 & 4 & 17\\
NiO & 2 & 14 & 9 \\
MgO  & 5 & 3 & 52\\
Al$_2$O$_3$ & 8 & 12 & 78 \\
Cr$_2$O$_3$ & 27 & 153 & 176\\
LiF & 3 & 18 & 20 \\
\end{tabular}
\end{center}
\end{table}


\begin{references}
\bibitem{PulayAdv} P. Pulay, Adv. Chem. Phys. {\bf 69}, 241 (1987).
\bibitem{PulayChapter} P. Pulay, in {\it Applications of Electronic Structure
Theory}, edited by H. F. Schaefer III, 153 (Plenum, New York, 1977).
\bibitem{Helgaker} T. Helgaker and P. J{\o}rgensen, Adv. in Quantum Chem.
{\bf 19}, 183 (1988).
\bibitem{HelgakerJorgensen1992} T. Helgaker and P. J{\o}rgensen,
in {\it Methods in Computational Physics}, edited by S. Wilson and
G. H. F. Diercksen, 353 (Plenum, New York, 1992).
\bibitem{SchlegelYarkony} H. B. Schlegel, in {\it Modern electronic 
structure theory}, edited by D. R. Yarkony, 459 (World Scientific, 
Singapore, 1995).
\bibitem{PulayYarkony} P. Pulay, in {\it Modern electronic 
structure theory}, edited by D. R. Yarkony, 1191 (World Scientific, 
Singapore, 1995).
\bibitem{Schlegel2000} H. B. Schlegel, Theor. Chem. Acc. {\bf 103},
294 (2000).
\bibitem{Pulay} P. Pulay, Mol. Phys. {\bf 17}, 197 (1969).
\bibitem{Bratoz} S. Brato\u{z}, in 
{\it Calcul des fonctions d'onde mol{\'e}culaire},
Colloq. Int. C. N. R. S. {\bf 82}, 287 (1958).
\bibitem{CRYSTALbuch} C. Pisani, R. Dovesi, and C. Roetti,
{\em Hartree-Fock Ab Initio Treatment of Crystalline Systems}, 
Lecture Notes in Chemistry Vol. 48 (Springer, Heidelberg, 1988).
\bibitem{Manual03} 
V. R. Saunders, R. Dovesi, C. Roetti, R. Orlando, C. M. Zicovich-Wilson ,
N. M. Harrison, K. Doll, B. Civalleri, I. J. Bush, Ph. D'Arco, M. Llunell,
{\sc crystal 2003} User's
Manual, University of Torino, Torino (2003).
\bibitem{IJQC} K. Doll, V. R. Saunders, N. M. Harrison,
Int. J. Quantum Chem. {\bf 82}, 1 (2001).
\bibitem{CPCarticle} K. Doll, Comput. Phys. Comm. {\bf 137}, 74 (2001).
\bibitem{MimmoArcoetal}B. Civalleri, Ph. D'Arco, R. Orlando, 
V. R. Saunders, R. Dovesi, Chem. Phys. Lett. {\bf 348}, 131 (2001).
\bibitem{Teramae198384} H. Teramae, T. Yamabe, C. Satoko and A. Imamura,
Chem. Phys. Lett. {\bf 101}, 149 (1983); H. Teramae, T. Yamabe and
A. Imamura, J. Chem. Phys. {\bf 81}, 3564 (1984).
\bibitem{Jacquemin} D. Jacquemin, J.-M. Andr{\'e} and B. Champagne,
J. Chem. Phys. {\bf 111}, 5306 (1999); J. Chem. Phys. {\bf 111}, 5324 (1999).
\bibitem{Hirata1997} S. Hirara and S. Iwata, J. Chem. Phys. {\bf 107},
10075 (1997).
\bibitem{Champagne2005} B. Champagne, in {\it Molecular Simulation
Methods for Predicting Polymer Properties}, edited by V. Galiasatos
(Wiley, New York, 2005)
\bibitem{Tobita2003} M. Tobita, S. Hirata, and R. J. Bartlett,
J. Chem. Phys. {\bf 118}, 5776 (2003).
\bibitem{Kudin2000} K. N. Kudin and G. E. Scuseria, Phys. Rev. B
{\bf 61}, 5141 (2000).
\bibitem{Kudin2001} K. N. Kudin, G. E. Scuseria, and H. B. Schlegel,
J. Phys. Chem. {\bf 114}, 2919 (2001).
\bibitem{Jacquemin2003} D. Jacquemin, J.-M. Andr\'e, and B. Champagne,
J. Chem. Phys. {\bf 118}, 373 (2003).
\bibitem{TCA2004} K. Doll, R. Dovesi and R. Orlando, Theor. Chem. Acc.
{\bf 112}, 394 (2004).
\bibitem{Ewald} P. P. Ewald, Ann. Phys. (Leipzig) {\bf 64}, 253 (1921).
\bibitem{VicCoulomb} V. R. Saunders, C. Freyria-Fava, R. Dovesi, L. Salasco,
and C. Roetti, Mol. Phys. {\bf 77}, 629 (1992).
\bibitem{Parry} D. E. Parry, Surf. Science {\bf 49}, 433 (1975);
{\bf 54}, 195 (1976) (Erratum).
\bibitem{Heyes} D. M. Heyes, M. Barber, and J. H. R. Clarke, 
J. Chem. Soc. Faraday Trans. II, {\bf 73}, 1485 (1977).
\bibitem{Vic1994} V. R. Saunders, C. Freyria-Fava, R. Dovesi, and C. Roetti,
Comp. Phys. Comm. {\bf 84}, 156 (1994).
\end{references}
\end{document}